\documentclass{iaus}
\usepackage{graphicx}

\title[Calibrating hydrogen-rich CC-SNe] 
{Calibrating hydrogen-rich core-collapse supernovae for their use as                                                                     
distance indicators independently of type Ia supernovae}

\author[Pumo \& Zampieri]   
{M.~L. Pumo$^{1,2,3}$ \and L. Zampieri$^1$
}

\affiliation{$^1$INAF-Osservatorio Astronomico di Padova, Vicolo dell'Osservatorio 5, I-35122 Padova, Italy 
\\[\affilskip]
$^2$Bonino-Pulejo Foundation, Via Uberto Bonino 15/C, I-98124 Messina, Italy
\\[\affilskip]
$^3$INAF-Osservatorio Astrofisico di Catania, via S. Sofia 78, I-95123 Catania, Italy
\\email: {\tt mlpumo@oact.inaf.it - luca.zampieri@oapd.inaf.it}}

\pubyear{2011}
\volume{281}  
\pagerange{119--126}
\jname{Binary Paths to Type Ia Supernovae Explosions}
\editors{R. Di Stefano \& M. Orio, eds.}
\begin{document}

\maketitle

\begin{abstract}
Using our new general-relativistic, radiation hydrodynamics, Lagrangian code, we
computed a rather extended grid of hydrogen-rich core-collapse supernova (CC-SN) models 
and explored the potentials of their ``standardization'' as distance indicators. We 
discuss the properties of some calibrations previously reported in the literature and 
present new correlations based on the behavior of the light curve, that can be employed 
for calibrating hydrogen-rich CC-SNe using only photometric data.
\keywords{supernovae: general, hydrodynamics, methods: numerical, distance scale}
\end{abstract}

\firstsection

\section{Introduction}

In recent years growing attention has been devoted to the construction of Hubble diagrams 
using hydrogen-rich CC-SN events (e.g., \cite[Olivares et al. 2010]{10} and references 
therein) in order to derive cosmological parameters independently of the usual method 
based on type Ia SNe (e.g., \cite[Freedman et al. 2009]{5} and references therein).
The possibility of building Hubble diagrams for hydrogen-rich CC-SNe is strictly related 
to the capability of calibrating them and, consequently, to turn them into usable distance 
indicators. Two different approaches are used to derive distance measurements of CC-SNe. One 
is based on theoretical spectral modelling like the expanding photosphere method (e.g., 
\cite[Eastman, Schmidt \& Kirshner 1996]{3}) or the ensuing spectral expanding atmosphere method 
(e.g., \cite[Baron et al. 2004]{2}). Others rely on more empirical techniques as the standardized 
candle method (e.g., \cite[Hamuy \& Pinto 2002, HP02]{6}), based on an observational correlation 
between the luminosity of a SN and its expansion velocity, or the method proposed by 
\cite[Elmhamdi, Chugai \& Danziger (2003, ECD03)]{4}, based on the steepness of the light curve.
In previous work we studied these correlations using a data/model comparison approach and tried to 
address their physical origin (e.g., \cite[Zampieri 2005, 2007]{15,16}). Along this vein, in the following 
we explore the existence of the correlations inferred from the aforementioned empirical methods in our 
model sample, which is composed of 22 models calculated with our new general-relativistic, radiation 
hydrodynamics, Lagrangian code (for details see \cite[Pumo, Zampieri \& Turatto 2010]{12} and 
\cite[Pumo \& Zampieri 2011]{11}, PZ11).

\section{Results and discussion}

\begin{figure}[!ht]
\centering
\includegraphics[angle=-90,width=1.74in]{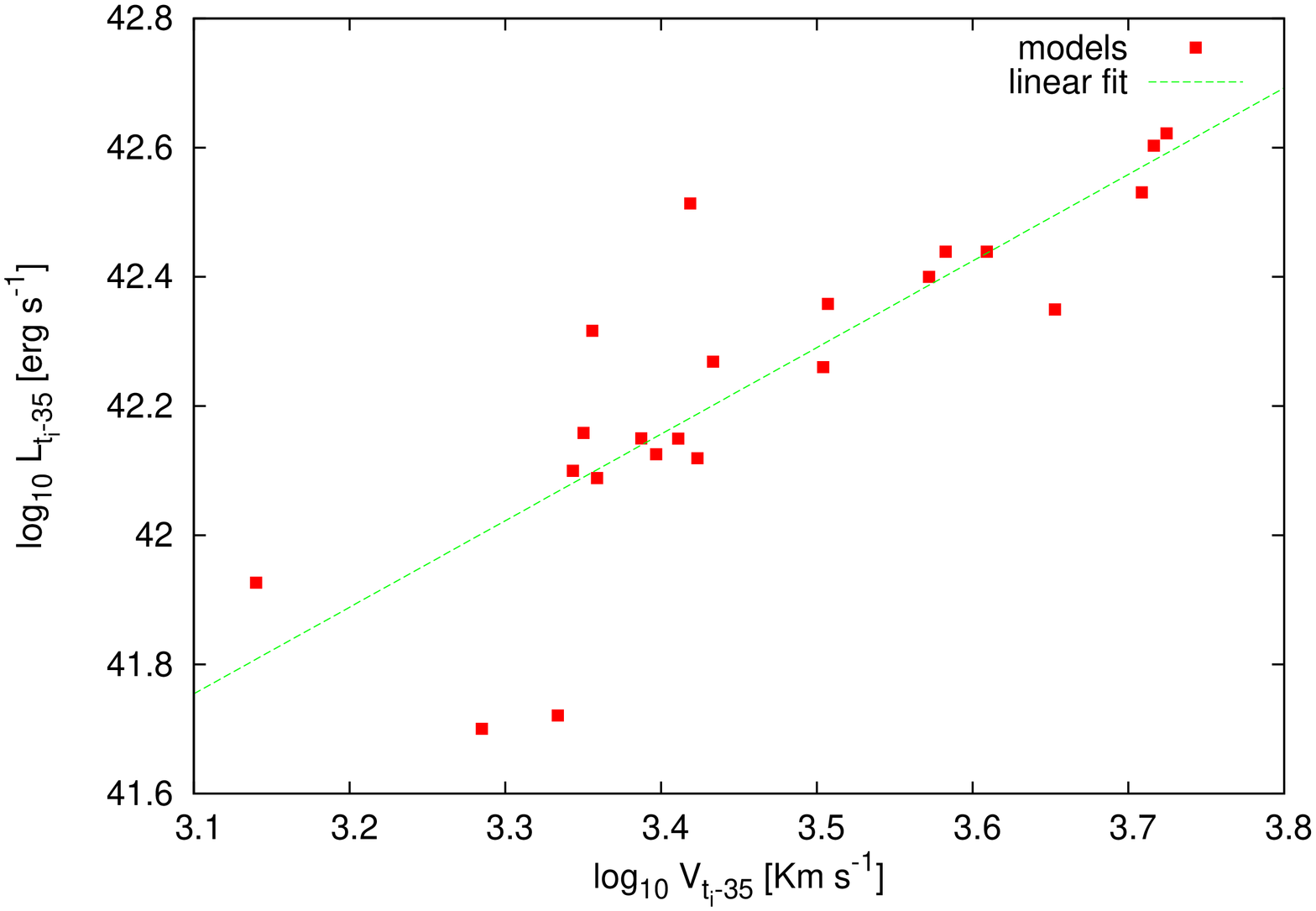}
\includegraphics[angle=-90,width=1.74in]{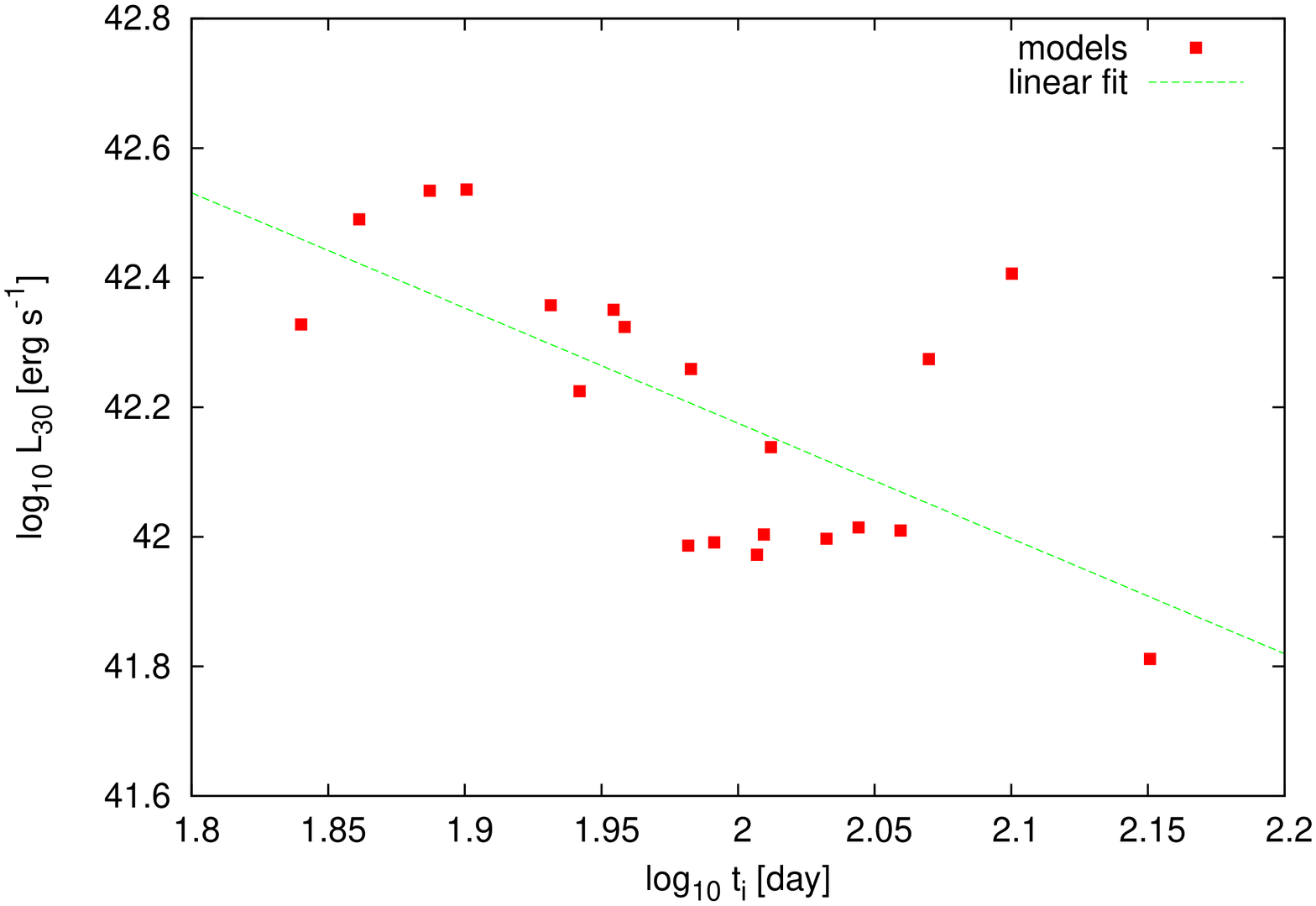}
\includegraphics[angle=-90,width=1.74in]{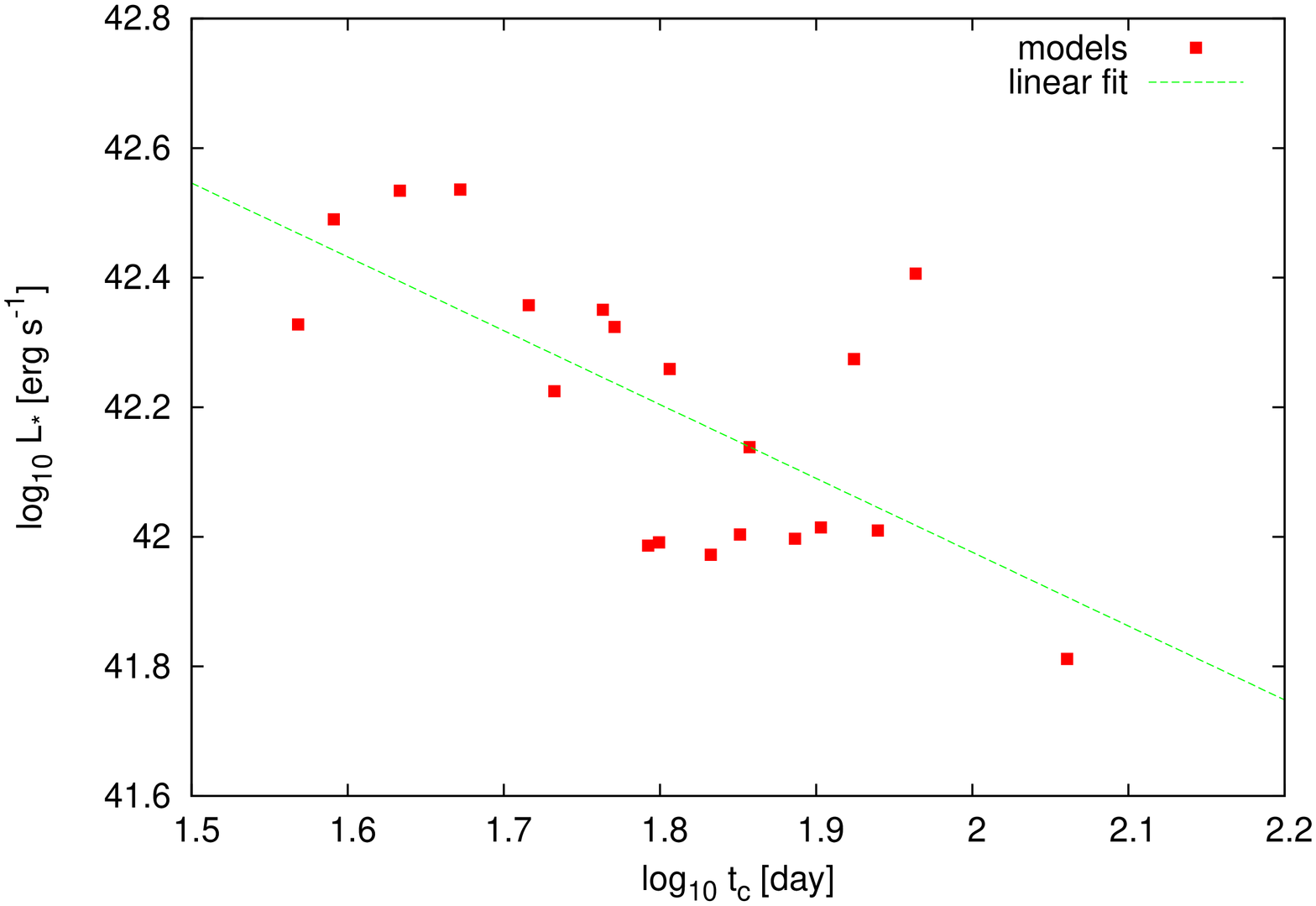}
\caption{{\it Left panel}: LEV relation for our 22 models, taking as reference epoch t$_i$-35 days. 
{\it Mid panel}: correlation between t$_i$ and L$_{30}$. {\it Right panel}: correlation between t$_c$ 
and L$_*$, taking as reference L$_*$=L$_{30}$. The correlations reported in mid and right panels are 
found excluding two SN 1987A-like models from the sample, retaining only models of type II plateau SNe.
The best linear fits of the three correlations are 
$\log_{10}\mathrm{L}_{t_i-35}= (1.34 \pm 0.21) \log_{10}\mathrm{V}_{t_i-35}$ $+$ $(37.60 \pm 0.72)$, 
$\log_{10}\mathrm{L}_{30}= - (1.78 \pm 0.48) \log_{10} t_i$ $+$ $(45.73 \pm 0.96)$, and 
$\log_{10}\mathrm{L}_*= - (1.14 \pm 0.29) \log_{10} t_c$ $+$ $(44.26 \pm 0.53)$, respectively. The null 
hypothesis probabilities inferred from the Pearson correlation coefficient are $< 10^{-3}$\%, 
$\lesssim 0.06$\%, and $\simeq 1.2$\%, respectively.}
\label{fig1}
\end{figure}

Our models reproduce the Luminosity-Expansion Velocity (LEV) relation (measured at 50 days from the 
explosion) with an index equal to 3.13 $\pm$ 0.29, in good agreement with the value 3.03 $\pm$ 0.37 
found in the observational sample of HP02. In order to overcome the difficulty of determining the phase 
at 50 days, that depends on the determination of the explosion time, we consider a LEV relation where 
the reference epoch is t$_i$-35 days (left panel in Fig.~\ref{fig1}), t$_i$ being the time when the 
semi-logarithmic derivative of the bolometric light curve S = -dlog$_{10}$L/dt has a local maximum at 
the end of the plateau phase. Moreover, our models confirm the anti-correlation between the light curve 
slope at t$_i$ and the amounts of $^{56}$Ni inferred by ECD03 on observational bases (see also PZ11), and 
show the existence of a relationship between t$_i$ and the bolometric luminosity L$_{30}$ on the plateau 
measured at 30 days from the explosion (mid panel in Fig.~\ref{fig1}). Furthermore, a preliminary analysis 
shows a very promising calibration relation between the luminosity L$_*$ at a generic time t$_*$ during the 
plateau and the characteristic time t$_c$=t$_{0.4}$-t$_*$, where t$_{0.4}$ is the time when L$_*$ decreases 
by a factor 2.5 (right panel in Fig.~\ref{fig1}).

The correlations shown in Fig.~\ref{fig1} represent useful tools for calibrating hydrogen-rich CC-SNe 
using only photometric data. In particular, for type II plateau SNe, the latter correlation is essentially 
independent of the explosion epoch (see also PZ11). 

While all the correlations reported here were obtained for a rather restricted sample of models, we are working 
at present to check their validity against a more extended grid of models, as well as against observations that 
are being collected within the ESO/TNG large program ``Supernova Variety and Nucleosynthesis Yields'' 
(PI: S.~Benetti).

\acknowledgments
M.L.P. acknowledges financial support from the Bonino-Pulejo Foundation.


\end{document}